\documentclass[runningheads]{llncs}
\usepackage[T1]{fontenc}
\usepackage{subcaption}
\usepackage{amsmath}
\usepackage{dirtytalk}
\usepackage{comment}
\usepackage{graphicx}
\usepackage{color}
\newcommand{\vf}[1]{\textcolor{blue}{\textbf{(VF: #1)}}}

\begin{document}
\title{Structure, Topics, and Diffusion Effects of Bluesky Starter Packs}
%
%
\author{
Andrea Failla\inst{1,2} \and
Vander L. S. Freitas\inst{3} \and
Giulio Rossetti\inst{1} \and
Carlos H. G. Ferreira\inst{3}
}

\authorrunning{A. Failla et al.}

\institute{
Institute of Information Science and Technologies “A. Faedo” (ISTI), 
National Research Council (CNR), Pisa, Italy\\
\email{\{name.surname\}@isti.cnr.it}
\and
Department of Computer Science, University of Pisa, Pisa, Italy
\and
Department of Computing, Universidade Federal de Ouro Preto (UFOP), Ouro Preto, MG, Brazil\\
\email{\{vander.freitas,chgferreira\}@ufop.edu.br}
}

%

%
\maketitle              
\begin{abstract}
User discovery is a central challenge in online social platforms, particularly during onboarding. Bluesky, a decentralized microblogging platform built on the AT Protocol, introduced starter packs: curated collections of accounts that users can follow in a single action to bootstrap their social network. In this paper, we present a large-scale empirical analysis of more than 50,000 English-language starter packs and over 600,000 associated users. We characterize their structural organization, topical composition, and impact on content diffusion. Our results show that starter packs form a highly interconnected ecosystem with substantial overlap across packs that largely reflects pre-existing communities. Topic modeling reveals a skewed landscape dominated by automatically generated personal packs alongside several thematic communities, which exhibit similar structural properties but markedly different adoption patterns. Finally, a matched event-study analysis shows that inclusion in a starter pack is strongly associated with a substantial increase in short-term repost activity.
\end{abstract}
\section{Introduction}
Online social platforms rely heavily on mechanisms that help users discover relevant accounts and communities~\cite{oubalahcen2023recommender,sharif2013review,gupta2013wtf}. As social networks grow in size and complexity, newcomers often face the \say{cold start} problem, as without existing connections it becomes difficult to identify interesting content and engage with the service~\cite{gupta2013wtf,zhang2024emergence}. To mitigate this issue, platforms have introduced a variety of discovery tools, including algorithmic recommenders and user-curated account lists~\cite{kang2012using,kywe2012survey,gupta2013wtf}.

Bluesky, a decentralized microblogging platform built on the Authenticated Transfer (AT) Protocol~\cite{kleppmann2024bluesky}, recently introduced \emph{starter packs}, i.e., curated collections of accounts and recommendation feeds that users can follow in a single action~\cite{Bluesky_2024_sp}.
Starter packs are designed to facilitate onboarding by enabling users to quickly bootstrap their social network. Prior work shows that they achieved rapid, large-scale adoption, driving up to 43\% of all daily follow operations at peak and playing a central role during major user influx events such as the 2024 U.S. elections~\cite{balduf2025bootstrapping}. Inclusion in a starter pack has also been shown to yield measurable benefits in terms of follower growth and content engagement~\cite{balduf2025bootstrapping}.

Despite this rapid adoption, our understanding of starter packs remains limited.
Existing work has primarily focused on their role in network growth and user-level benefits~\cite{balduf2025bootstrapping,smith2026blue}, leaving several fundamental aspects unexplored.
First, the structural organization of starter packs as a system remains unclear, as it is unknown whether they form isolated collections or a broader, interconnected ecosystem, and what follower network topologies emerge among their members. Second, the topical landscape of starter packs is not well characterized, since prior analyses provide only coarse categorizations without examining how community type relates to structural properties and adoption dynamics. Finally, the impact of starter pack inclusion on content diffusion remains an open question, as prior work considers followers and likes as proxies for visibility, but it is unclear whether these mechanisms translate into increased propagation of content.

This paper addresses these gaps through a large-scale empirical analysis of starter packs on Bluesky. 
We aim to answer the following research questions: (RQ1.) What structural properties characterize Bluesky starter packs and the follower networks they induce?
(RQ2.) How do starter packs vary across topics in terms of composition, community structure, and adoption?
(RQ3.) Is inclusion in a starter pack associated with changes in the diffusion of users’ content?
Analyzing more than 50K starter packs, we find that packs show substantial user overlap and dense, reciprocal follower networks that largely reflect pre-existing communities. Topic modeling reveals a skewed landscape dominated by automatically generated personal packs, alongside a smaller set of well-defined thematic communities with heterogeneous adoption. Finally, inclusion in a starter pack is associated with a significant short-term increase in content diffusion, with higher repost activity relative to matched controls.

The rest of the paper is organized as follows. 
Section~\ref{sec:background} provides background on Bluesky and related work on social discovery mechanisms. 
Section~3 describes the dataset and methodological framework used in our analysis. 
Section~4 presents the empirical results on the structure, topics, and diffusion effects of starter packs. 
Finally, Section~5 concludes and discusses limitations and directions for future work.

\section{Background and Related Work}\label{sec:background}
In the following, we introduce a primer on the Bluesky platform and provide context on social discovery on online social media.
\\ \ \\
\noindent\textbf{Bluesky Social.} Bluesky\footnote{\url{https://bsky.app/}} is a decentralized microblogging platform that supports short-form posts and social interactions similar to established services such as X (formerly Twitter). The platform is built on the Authenticated Transfer (AT) Protocol, which enables a federated and extensible social networking ecosystem~\cite{kleppmann2024bluesky}. User data is stored in repositories hosted on \textit{Personal Data Servers} (PDS), which maintain signed records of user activity and expose public endpoints that can be queried to retrieve the complete user history. 
These design choices make Bluesky particularly suitable for large-scale empirical research, as user activity and network structure can be systematically observed through public interfaces.

Bluesky also enables \emph{algorithmic choice}, allowing users to subscribe to \textit{feed generators}, i.e., custom recommendation algorithms that filter or rank posts according to specific criteria and can be created and shared by users~\cite{failla2024m}. In addition, Bluesky introduced \emph{starter packs}, curated collections of accounts and feeds that users can follow in a single action to bootstrap their social network~\cite{Bluesky_2024_sp,smith2026blue,balduf2025bootstrapping}.
New users can also sign up directly through a starter pack, which automatically initializes their social graph by following its members. For registered users, the platform offers a \say{Make One for Me} feature, to automatically build a pre-populated starter pack of suggested users and custom feeds~\cite{Bluesky_2024_sp}, though the algorithmic rationale behind these suggestions remains opaque.
\\ \ \\
\noindent\textbf{Social Discovery.} The problem of recommending other users/content to a social media user is often framed as a link prediction problem, where the goal is to infer which connections are likely to form in the future based on observed network structure and user behavior~\cite{liben2003link}. 
For example, Twitter (X) developed the \say{Who to Follow} recommendation system, which suggests accounts based on graph-based algorithms using shared connections, interactions, and interest similarity~\cite{gupta2013wtf}. 
Similar recommendation systems have been deployed on other large-scale social platforms. 
For instance, LinkedIn's \say{People You May Know} feature employs a large-scale recommendation pipeline that combines graph-based candidate generation (e.g., random walks over the social graph) with learned user representations to identify and rank potential connections~\cite{linkedinBuildingLargeScale}.
In addition to algorithmic approaches, platforms have also adopted community-driven discovery mechanisms.
Examples include Reddit's topic-based communities (subreddits), which organize users and content around shared interests~\cite{soliman2019characterization}, and Twitter Lists, which allow users to curate collections of accounts around specific themes and communities~\cite{kang2012using}.
More recently, some platforms have attempted to mitigate the user discovery problem during onboarding by importing a user's existing social graph from other services. For instance, Meta's Threads initially allowed new users to automatically follow accounts they already followed on Instagram, partially circumventing the cold-start problem~\cite{zhang2024emergence,kim2024does}. 
These approaches highlight two complementary paradigms for social discovery, namely algorithmic recommendation and community-driven curation. 
Starter packs can be seen as a hybrid mechanism that combines elements of both.
\\ \ \\
\noindent\textbf{Starter Packs in Prior Work.}
The literature explicitly focusing on the Starter Pack mechanism remains limited. 
Existing work has primarily focused on their role in platform growth and user-level outcomes.
In particular, Balduf et al.~\cite{balduf2025bootstrapping} show that starter packs can drive a substantial fraction of follow activity and yield measurable benefits in terms of followers, likes, and user engagement, while Smith et al.~\cite{smith2026blue} provide a large-scale dataset that enables the study of starter packs as higher-order social structures. 
Beyond these contributions, prior research on social discovery and the Bluesky platform has largely focused on network structure, and platform-level dynamics~\cite{quelle2025bluesky,failla2024m}, without specifically addressing starter packs. 
As a result, several fundamental questions remain open, including how starter packs are structurally organized as an interconnected system, how their topical composition reflects different communities, and whether their use affects the diffusion of content across the network.

\section{Materials and Methods}
In this section, we outline the methodological tools used in this work.
Code to reproduce the figures and statistics is available on Github.\footnote{\url{https://github.com/andreafailla/bsk-sp-asonam2026}}
\subsection{Data Collection}
Starting from a $\sim 5$M Bluesky user sample~\cite{failla2024m}, we query the platform API endpoints to obtain user PDSes containing activity up to February 2025. 
This updated dataset includes posts with associated metadata, follow relationships, blocks, likes, feed generator subscriptions, and starter packs.
Focusing on starter packs, we identify $81$K starter packs created by users in our initial sample.
Each starter pack has a title and a description, and is associated with a list of accounts and (optionally) feed generators.
To focus our analysis on a linguistically homogeneous subset, we detect the language of the starter pack descriptions using Google's \texttt{langdetect}\footnote{\url{https://pypi.org/project/langdetect/}} library.
Overall, we identify approximately $50$K ($\sim 60\%$)English starter packs populated by $666$K unique users.
Among these users, $293$K were not part of our original sample.
For completeness, we therefore collect their public activity and metadata as well, expanding the coverage of our dataset. The remainder of the study focuses exclusively on the $50$K English starter packs and their associated users.

\subsection{Network Modeling}

To study the structure of starter packs, we construct a co-occurrence network based on starter pack membership.
We model a bipartite network $B=(U,S,E)$ where $U$ represents users, $S$ represents starter packs, and edges in $E$ indicate that a user is included in a given starter pack.
We then project this bipartite network onto the starter pack layer to obtain a weighted starter-pack co-occurrence network.
In this projection, two starter packs are connected if they share at least one user, and the weight $w_{ij}$ of the edge corresponds to the number of users appearing in both packs. 
\\ \ \\
Because projections of bipartite networks tend to be dense and noisy, obscuring the underlying structural organization of the system~\cite{yassin2023evaluation,gomes2022network,serrano2009extracting}, we extract the statistically significant backbone of the network using the Disparity Filter algorithm~\cite{serrano2009extracting}. For a node $i$ with degree $k_i$ and strength $s_i=\sum_j w_{ij}$, the normalized weight of edge $(i,j)$ is $p_{ij}=w_{ij}/s_i$. Under the null hypothesis that the strength $s_i$ is uniformly distributed across its $k_i$ edges, the probability that a normalized weight is compatible with random allocation is $\alpha_{ij}=1-(k_i-1)\int_0^{p_{ij}} (1-x)^{k_i-2}\,dx$. Edges are retained in the backbone if $\alpha_{ij}<\alpha$, where $\alpha$ is a chosen significance threshold. Applying the disparity filter yields a sparse backbone network that preserves the most informative relationships between starter packs while removing statistically insignificant connections.
\\ \ \\
Finally, we associate each starter pack $s \in S$ with the follower network induced by its members, denoted $G_s=(V_s,E_s)$, where $V_s$ is the set of users in $s$ and $E_s$ represents the directed following relationships among them.
This induced subgraph captures the internal connectivity of the starter pack. 
Importantly, these induced networks are not necessarily cliques: membership in the same starter pack does not imply that all users follow each other. Instead, the induced subgraph captures the subset of \textit{pre-existing follower ties} among pack members.

\subsection{Topic Extraction}
To characterize the thematic structure of starter packs, we extract topics from their textual descriptions using \textit{BERTopic}~\cite{grootendorst2022bertopic}.  BERTopic is a topic modeling framework that leverages transformer-based embeddings to capture semantic similarities between documents. 
Specifically, we first encode the text of each starter pack description using a pre-trained sentence transformer model, obtaining dense vector representations of the documents. These embeddings are then reduced in dimensionality using Uniform Manifold Approximation and Projection for Dimension Reduction (UMAP), which preserves local semantic structure while facilitating clustering~\cite{mcinnes2018umap}.
Next, we apply the \textit{HDBSCAN} clustering algorithm to group semantically similar descriptions into clusters corresponding to latent topics~\cite{mcinnes2017hdbscan}. Unless otherwise specified, we adopt the default parameterization recommended in the BERTopic documentation.
For each cluster, BERTopic identifies representative keywords using a class-based term frequency--inverse document frequency (cTF-IDF) weighting scheme, which highlights terms that are particularly distinctive for that topic relative to the corpus. Finally, we leverage \textit{GPT-4.5}~\cite{achiam2023gpt} to improve the interpretability of the extracted topics, following the LLM-based topic representation approach proposed in the BERTopic framework. Specifically, we use the prompt recommended in the official documentation, which provides the model with the most representative documents and keywords for each topic and asks it to generate a concise topic label.\\footnote{\url{https://maartengr.github.io/BERTopic/getting\_started/representation/llm.html}}. 
For each topic, we provide the model with the most representative keywords and a sample of documents from the cluster to generate a concise name and description. We also manually review 50 random documents per topic to ensure label quality and consistency.

\subsection{Assessing the Impact of Starter Packs on Post Performance}
Finally, we evaluate whether first-time inclusion in a starter pack is associated with changes in post diffusion using a matched event-study design at the author level to address our third objective.
Specifically, we test whether being included in a starter pack affects content diffusion in terms of reposts.
\\ \ \\
\noindent\textbf{Treatment and Outcomes.} For each author, the treatment time is defined as the first observed moment at which the author is added to any starter pack. We then link posts and repost events to authors and construct post-level diffusion outcomes measured within a seven-day window after publication. Specifically, for each post we record (i) whether the post receives at least one repost within seven days and (ii) the total number of reposts accumulated during that period. This allows us to distinguish between the likelihood of diffusion and its magnitude. Post-level outcomes are subsequently aggregated to the author--week level, producing a panel in which each observation records the number of posts published by an author in a given week, the number of those posts that receive at least one repost, and the total number of reposts obtained.
\\ \ \\
\textbf{Matching.} For treated authors, weeks are aligned in event time relative to the week of first inclusion in a starter pack, such that \(t=0\) denotes the inclusion week. This re-indexing produces an event-time panel that allows us to track repost outcomes before and after treatment. Authors who are never included in any starter pack serve as potential controls. To improve comparability, each treated author is matched to \(k=5\) never-treated authors using nearest-neighbor matching with exact matching on pre-treatment follower-count deciles. 
Within each follower-count stratum, distance is computed using pre-treatment repost outcomes, namely the share of posts that receive at least one repost, the average number of reposts per post, and the corresponding linear pre-treatment trends. Follower counts are measured strictly before the treatment week for treated authors and at an analogous pre-period anchor week for controls, ensuring that matching is based only on information available prior to treatment.
Matched control authors are then used to construct counterfactual outcomes in the same calendar weeks as treated observations, allowing us to compare treated and matched-control repost outcomes both before and after inclusion.
\\ \ \\
\textbf{Difference-in-Differences Event Study.} We estimate the dynamic effect of starter-pack inclusion using a matched event-study difference-in-differences framework.
For each event week \(\tau\), we calculate the mean outcome among treated authors observed at that relative time and subtract the weighted mean outcome of their matched controls observed in the same calendar weeks.
Control authors are reweighted according to the number of times they are selected as matches across treated authors. Formally, for each outcome \(Y\), we define: $\hat{\beta}_{\tau} = \bar{Y}^{\text{treated}}_{\tau} - \bar{Y}^{\text{control}}_{\tau}$,
where \(\bar{Y}^{\text{treated}}_{\tau}\) is the treated mean outcome at event time \(\tau\), and \(\bar{Y}^{\text{control}}_{\tau}\) is the corresponding weighted matched-control mean in the same calendar weeks.
To summarize the dynamics, we compare a pre-treatment window \((\tau=-8,\ldots,-1)\) with a post-treatment window \((\tau=0,\ldots,8)\).
The overall difference-in-differences estimate is defined as the change in the treated-control gap from the pre-treatment to the post-treatment period, that is, the post-period matched difference minus the pre-period matched difference.
Finally, a key assumption of the design is that, in the absence of treatment, treated and control authors would have followed parallel trends. To assess the plausibility of this assumption, we examine pre-treatment differences in outcomes using the event-study estimates. Specifically, we compute the matched treated-control differences $\hat{\beta}_\tau$ for pre-treatment periods $\tau < 0$ and inspect whether these coefficients are close to zero and stable over time.

\section{Results}

\begin{figure}[t]
    \centering

    \begin{subfigure}{0.48\linewidth}
        \centering
        \includegraphics[width=\linewidth]{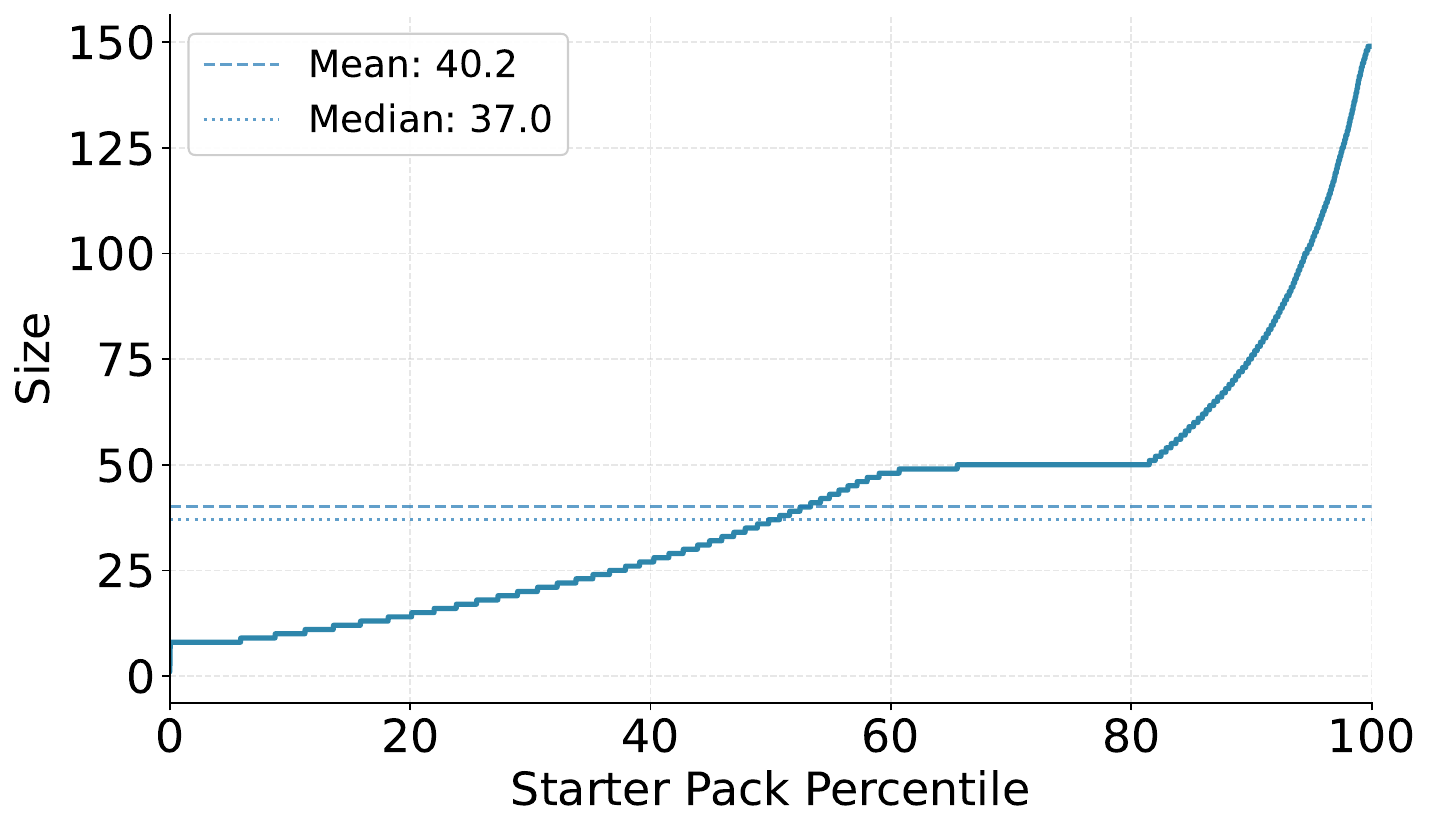}
        \caption{}
        \label{fig:sizes}
    \end{subfigure}
    \hfill
    \begin{subfigure}{0.48\linewidth}
        \centering
        \includegraphics[width=\linewidth]{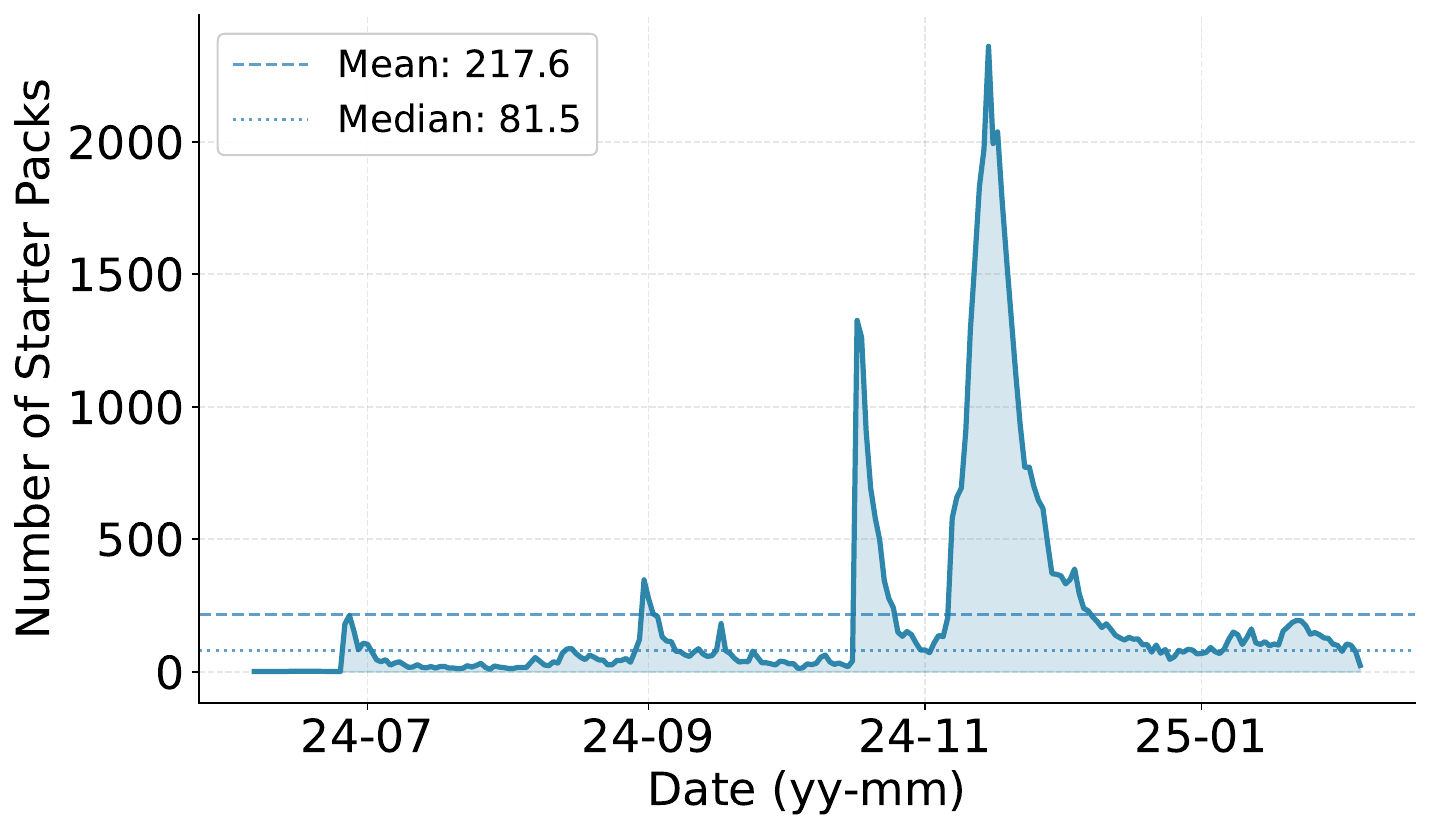}
        \caption{}
        \label{fig:times}
    \end{subfigure}

    \caption{Starter pack dataset statistics. Panel (a) shows the cumulative distribution of starter pack sizes. Panel (b) shows the number of starter packs created each day. Dashed lines indicate mean values, while dotted lines indicate median values.}
    \label{fig:combined}
\end{figure}

\noindent \textbf{Starter Pack Overview.} 
Starter packs are generally moderate in size (median=37, mean=40.2 users), indicating that most packs contain a few dozen accounts (Fig.~\ref{fig:sizes}). 
Pack sizes increase gradually across percentiles, suggesting a relatively smooth distribution without sharp concentration points.
A noticeable plateau appears around 50 users, where many packs share the same size. 
This is consistent with the platform’s automatic generation feature (\say{Make one for me}), which creates packs of 50 users. 
Beyond roughly the 80th percentile, pack sizes grow more rapidly, with the largest packs approaching the platform limit of 150 users.
We observe that starter pack creation is highly uneven across users. 
While most users created only a single pack, a small number of accounts produced a large number of starter packs. 
For example, the most active creator generated 107 packs, followed by others with 68, 60, and 52 packs, respectively (not shown). 
A manual inspection of these highly prolific creators reveals that many of their packs are very large (close to the platform limit of $\sim150$ members) and follow systematic naming conventions such as ``Democrats on Bluesky \#1'', ``Democrats on Bluesky \#2'', and so on. 
This pattern indicates that some users may have generated multiple packs potentially to organize large communities and/or to circumvent the platform’s maximum pack size limitation.
Fig.~\ref{fig:times} shows the temporal distribution of starter pack creation, measured as the number of packs generated per day. 
Activity remained relatively low in the first months following the feature’s launch in June 2024. 
However, creation rates increase sharply beginning in October-November 2024, with several large spikes.
These spikes coincide with documented waves of user migration to Bluesky. 
In October 2024, changes to policies on X (Twitter), including updates to the block feature and terms related to AI training on user content, triggered a migration of users to alternative platforms.
Shortly afterward, Bluesky experienced rapid growth, gaining over one million new users within a few days. 
A second major surge occurred around November 2024, following the 2024 United States presidential election, when Bluesky’s user base expanded rapidly and surpassed 20 million users within weeks. 
The influx of newcomers likely increased the demand for curated account lists, leading to a surge in starter pack creation.
 \\ \ \\
\textbf{Overlap Network Analysis.} Table~\ref{tab:graph_properties} reports summary statistics of the starter-pack overlap network. 
The original graph $\mathcal{G}$ is highly interconnected, indicating that many starter packs share users. 
With a density of $0.0736$ and an average degree of $3703.36$, each pack overlaps with thousands of others on average, reflecting substantial redundancy in the sets of curated users. 
The network is also almost fully connected: only $8$ components exist, and the largest component contains $99.97\%$ of all nodes, showing that most packs are indirectly linked through shared users.
These patterns indicate that starter packs are far from independent curated lists. 
In other words, different packs often represent overlapping views of a shared pool of visible accounts rather than distinct curatorial efforts.
Applying the disparity filter produces a much sparser backbone network $\mathcal{G}^{df}$ by retaining only statistically significant overlaps.
Specifically, we retain edges whose significance level satisfies $\alpha < 0.05$, following the standard threshold used in prior work on backbone extraction~\cite{serrano2009extracting,gomes2022network}.
This filtering step removes approximately 96\% of edges and reduces the number of nodes by about 23\%. 
The resulting backbone network remains largely connected: although the number of components increases to 684, the largest component still contains 95.83\% of nodes.
The persistence of a large connected component even after aggressive filtering suggests that the overlap between packs is not purely random. 
Instead, it reflects systematic patterns of user co-occurrence, likely driven by shared topical communities and/or by repeated inclusion of prominent accounts across multiple lists.

\begin{table}[h]
    \centering
    \tiny
        \caption{Starter pack overlap network properties}
    \label{tab:graph_properties}
    \begin{tabular}{lcccccc}
        \hline
         & $|V|$ & $|E|$  & Avg. Degree (std) & Density & Components & Largest CC (\%) \\
        \hline
        $\mathcal{G}$ & 50289 & 93119111 & 3703.36 (4986.62) & 0.0736 & 8 & 99.97 \\
        $\mathcal{G}^{df}$ & 38546 & 3187359 & 165.38 (420.44) & 0.00429 & 684 & 95.83 \\
        \hline
    \end{tabular}

\end{table}

\begin{figure}[t]
    \centering

    \begin{subfigure}{0.48\linewidth}
        \centering
        \includegraphics[width=\linewidth]{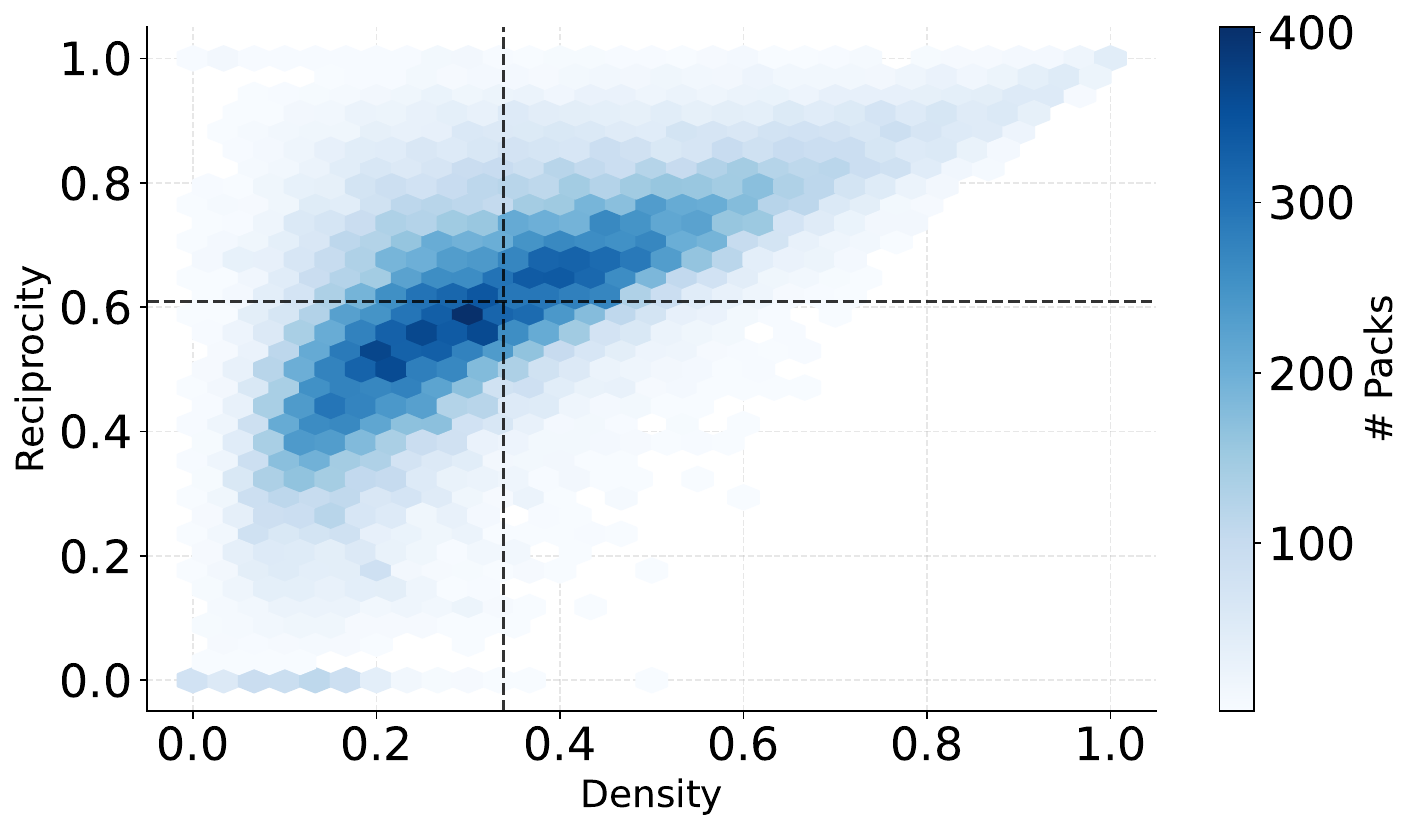}
        \caption{}
        \label{fig:reciprociry}
    \end{subfigure}
    \hfill
    \begin{subfigure}{0.48\linewidth}
        \centering
        \includegraphics[width=\linewidth]{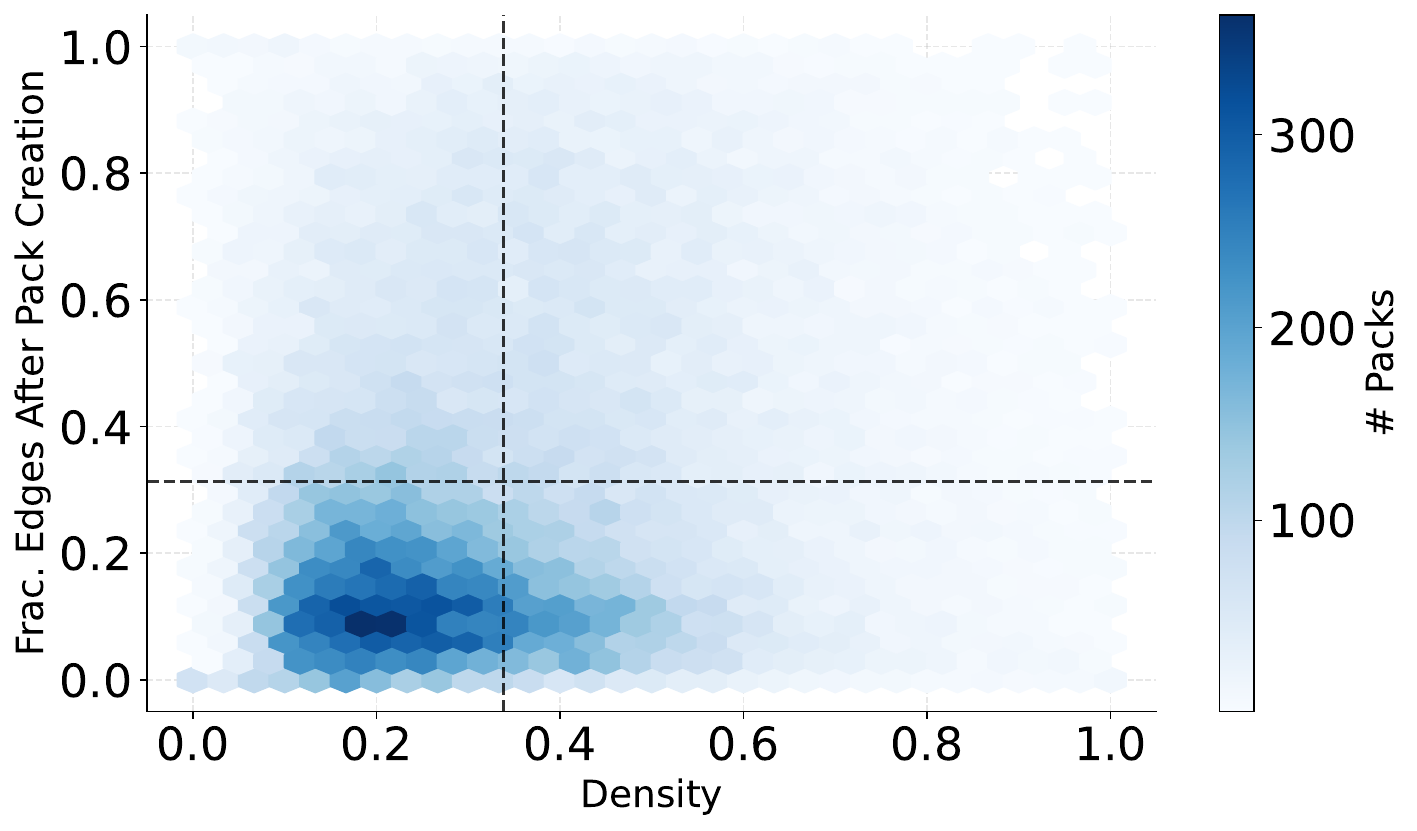}
        \caption{}
        \label{fig:fraction}
    \end{subfigure}

    \caption{Reciprocity (a) and Fraction of edges created after pack creation (b) as a function of network density for starter pack follower networks $G_s$. 
    Dashed lines indicate average values.}
    \label{fig:combined1}
\end{figure}

\noindent\textbf{Starter Pack Follower Networks.}
Across the dataset, most starter packs correspond to internally connected follower networks. Specifically, 46,516 out of 51,422 packs (90\%) form a single connected component when considering follower relationships among their members. On average, these induced networks exhibit a density of 0.338, indicating that a substantial fraction of possible connections between members are present. Reciprocity is also relatively high (mean $0.608$), thus many follower relationships among pack members are mutual.
Fig.~\ref{fig:reciprociry} highlights that follower networks cluster around these average density and reciprocity values, but also show a concentration toward highly dense and highly reciprocal networks, with substantially wider variance at lower density.
\\ 
We further examine the temporal dynamics of these connections. On average, 31.4\% of the edges within each starter pack network were created after the pack itself was formed. This indicates that while many ties among members already existed before pack creation, starter packs do not necessarily correspond to newly formed communities.
\\
Fig.~\ref{fig:fraction} shows that most starter pack networks exhibit relatively low fractions of post-creation edges, with the highest concentration below 0.2. In other words, the majority of connections among pack members predate the pack itself. 
At the same time, lower-density packs display substantially greater variability, suggesting that starter packs can occasionally facilitate the formation of new ties among users who were not previously connected.
\\
To further investigate this mechanism, we approximate direct starter-pack adoption events. Because following a starter pack results in multiple nearly simultaneous follow actions~\cite{balduf2025bootstrapping}, we identify bursts of follow activity as potential starter-pack adoptions. 
Specifically, for a given user and starter pack, we classify a \textit{starter-pack adoption event} when at least 50\% of the pack's members are followed within a 5-second time window. 
Using this heuristic, we identify more than 100K such events, involving approximately 61K distinct members, i.e., about $9\%$ of all users appearing in starter packs.
Overall, these results suggest that while starter packs are sometimes used as discovery mechanisms by external users, their direct effect on the network formation of their members appears relatively limited. 
A possible explanation is that users are not notified when they are added to a starter pack, thereby reducing the incentive to engage actively with other members.
Consequently, starter packs primarily surface existing communities, with most follow events occurring by outsiders. 
It should be noted, however, that since our heuristic cannot detect cases in which users already follow some or all pack members before the burst, these estimates should be interpreted as a lower bound on starter-pack follow events.
\\ \ \\
\noindent\textbf{Topic Analysis.} To further understand how these structural patterns relate to the content represented by starter packs, we analyze their topical composition using BERTopic. We generate text embeddings with the \texttt{all-MiniLM-L6-v2}\footnote{\url{https://huggingface.co/sentence-transformers/all-MiniLM-L6-v2}} sentence-transformer model and reduce dimensions via UMAP ($n\_neighbors=3$, $n\_components=5$, $\text{min\_dist}=0.1$, cosine metric), followed by HDBSCAN clustering ($\text{min\_cluster\_size}=100$). 
We then perform TF-IDF-based outlier reduction\footnote{\tiny\url{https://maartengr.github.io/BERTopic/getting\_started/outlier\_reduction/outlier\_reduction.html}} so that all data points are assigned a topic. This procedure yields 11 distinct topics, summarized in Table~\ref{tab:sp_topics}.

The distribution of topics is highly skewed as the majority of starter packs fall into the category of personal packs (``[username]'s starterpack''), which account for more than half of all packs in the dataset.
These packs are typically generated through Bluesky's built-in ``Make one for me'' feature, which automatically assembles a set of accounts. 
As a result, many starter packs reflect algorithmically suggested communities rather than manually curated thematic groupings.
Beyond these personal collections, several recurring thematic communities emerge. 
Creative and knowledge-oriented communities, such as \textit{Art and Writing} and \textit{Science and Research}, represent some of the largest groups. 
Other prominent themes include \textit{Gaming}, \textit{News and Journalism}, and \textit{Content Creators}, reflecting the presence of established online communities on the platform. 
In addition, we identify several smaller but clearly defined niche topics, including \textit{Tabletop RPGs}, \textit{Cats}, and discussions related to the \textit{Ukraine and Russia} conflict.
Finally, a distinct category of starter packs relates to platform migration. 
These packs explicitly reference users transitioning from Twitter, highlighting the role of starter packs as a tool for rebuilding social networks when joining Bluesky.
\begin{table}[h!]
\centering
\tiny
\caption{Summary of detected starter pack Topics.}
\label{tab:sp_topics}
\resizebox{\linewidth}{!}{
\begin{tabular}{|l|p{8cm}|c|}
\hline
\textbf{Topic} & \textbf{Description} & \textbf{\#SPs} \\
\hline
People ([username]'s starterpack) & Personal packs. This is the default name for automatically-created packs. & 29874 \\
\hline
Art and Writing & Communities focused on sharing artwork, creative processes, literature, poetry, and writing tips. & 7635 \\
\hline
Science and Research & Packs connecting scientists, researchers, science communicators, and enthusiasts across various disciplines. & 4589 \\
\hline
Gaming & Packs related to video gaming communities, covering various gaming platforms, genres, esports, and gaming culture. & 3459 \\
\hline
News and Journalism & Starter packs aggregating journalists, news media, breaking news updates, investigative journalism, and media literacy. & 2183 \\
\hline
Content Creators & Groups assembling influencers, streamers, podcasters, YouTubers, and other digital media creators. & 2159 \\
\hline
Twitter Migration & Starter packs aimed at users migrating from Twitter, focused on rebuilding networks and maintaining online communities post-migration. & 442 \\
\hline
Comedy and Laughter & Groups centered around humor, comedy content, jokes, memes, and comedic social interactions. & 303 \\
\hline
Tabletop RPGs & Communities interested in tabletop role-playing games, sharing gameplay stories, resources, game mechanics, and player groups. & 289 \\
\hline
Cats & Packs dedicated to cat enthusiasts, sharing cat photos, videos, memes, and pet-care advice. & 247 \\
\hline
Ukraine and Russia & Communities and resources dedicated to the discussion of geopolitical, social, humanitarian, and informational aspects of the Russia-Ukraine conflict. & 242\\\hline
\end{tabular}
}
\end{table}

\begin{figure}
    \centering
    \includegraphics[width=.8\linewidth]{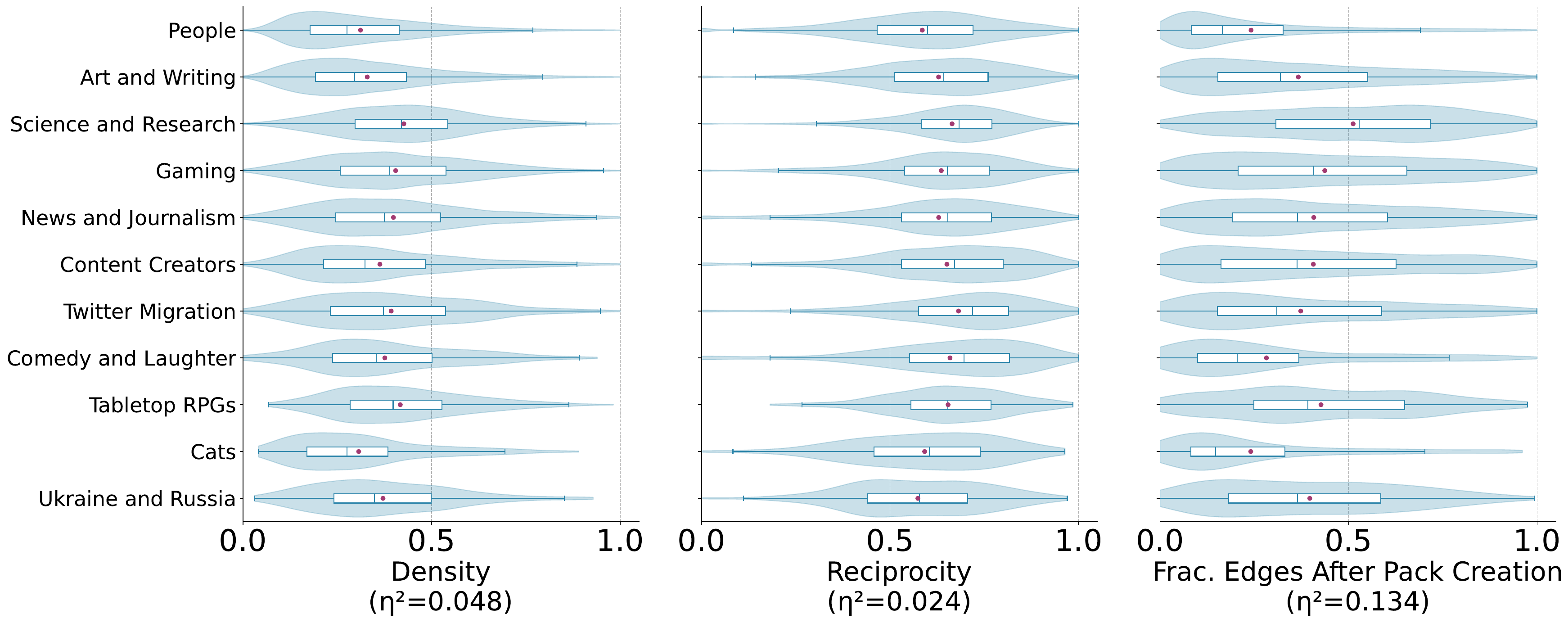}
    \caption{Distribution of follower network structure metrics across topics.
    Violin plots show the full distribution of values for each topic, with box plots indicating the interquartile range and median. 
    The red dots mark the mean of each distribution. $\eta^2$ measures the proportion of variance explained by topic membership.}
    \label{fig:follownets}
\end{figure}
\begin{figure}
    \centering
    \includegraphics[width=.7\linewidth]{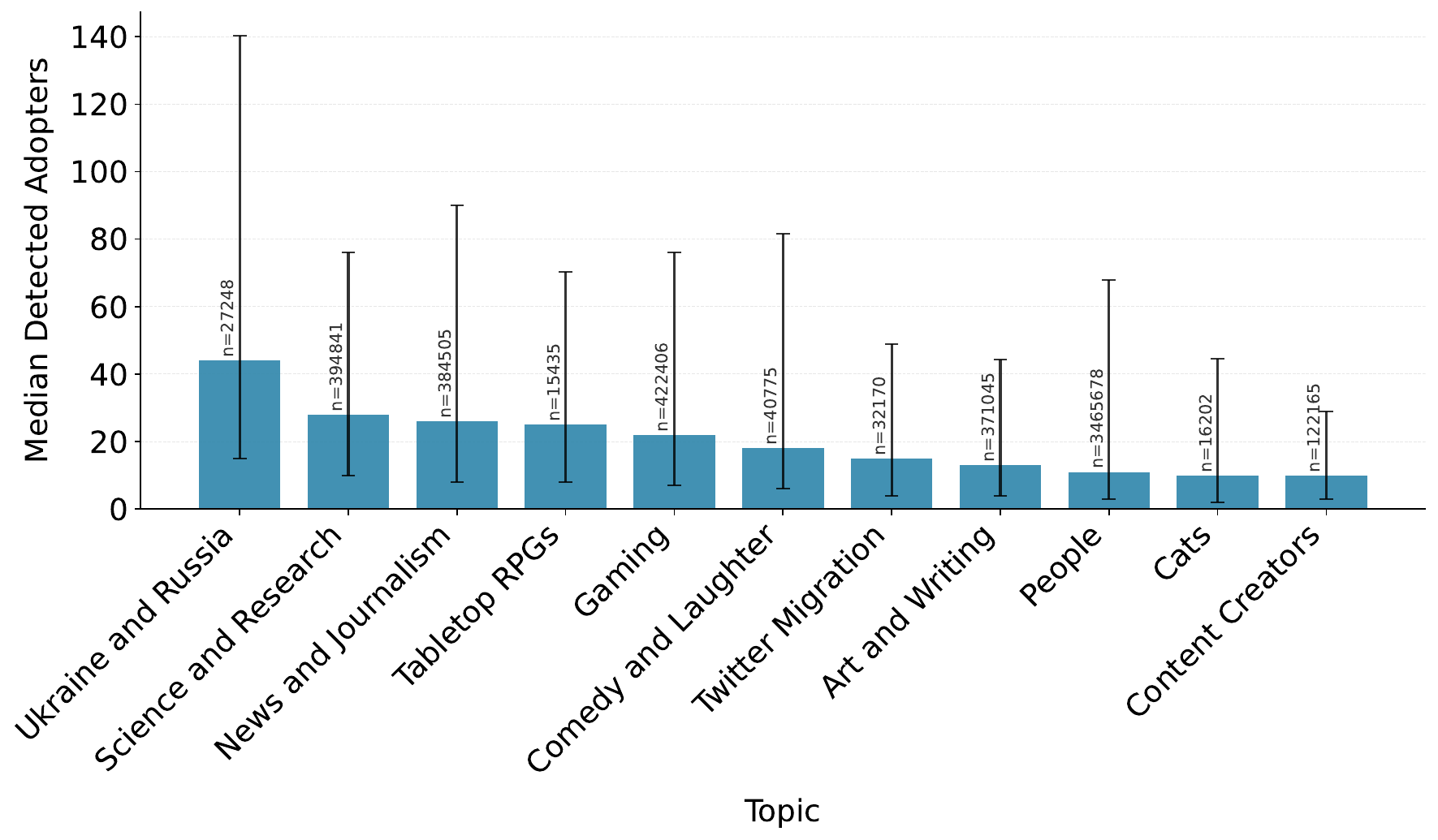}
    \caption{Median starter-pack adoption by topic. Bars show the median number of detected adopters per starter pack within each topic, with error bars indicating the interquartile range. $n$ denotes the total number of detected adopters aggregated across all starter packs in the corresponding topic. }
    \label{fig:adoption}
\end{figure}
\begin{figure}[t]
    \centering

    \begin{subfigure}{0.48\linewidth}
        \centering
        \includegraphics[width=\linewidth]{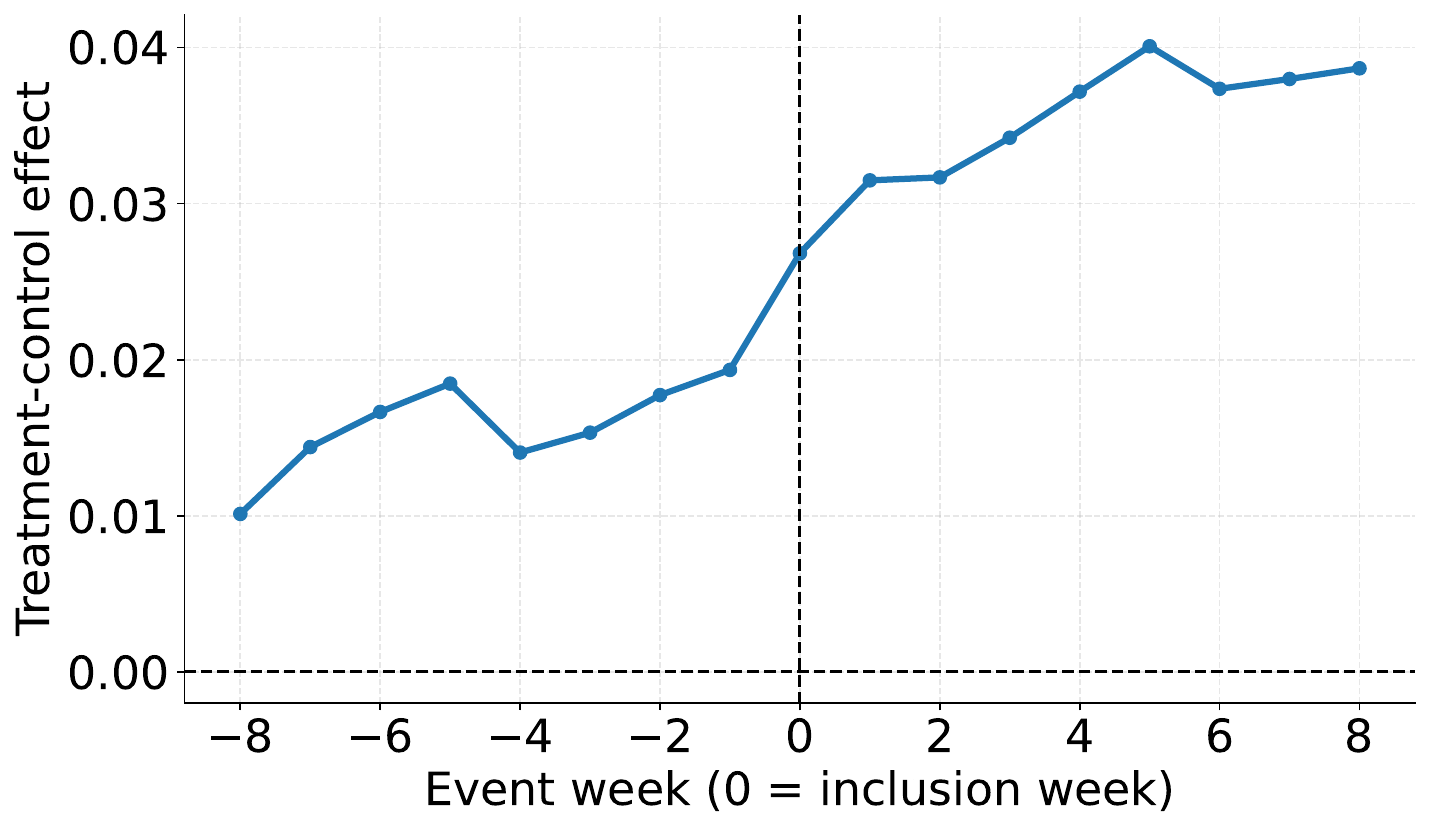}
        \caption{}
        \label{fig:pretrend-once}
    \end{subfigure}
    \hfill
    \begin{subfigure}{0.48\linewidth}
        \centering
        \includegraphics[width=\linewidth]{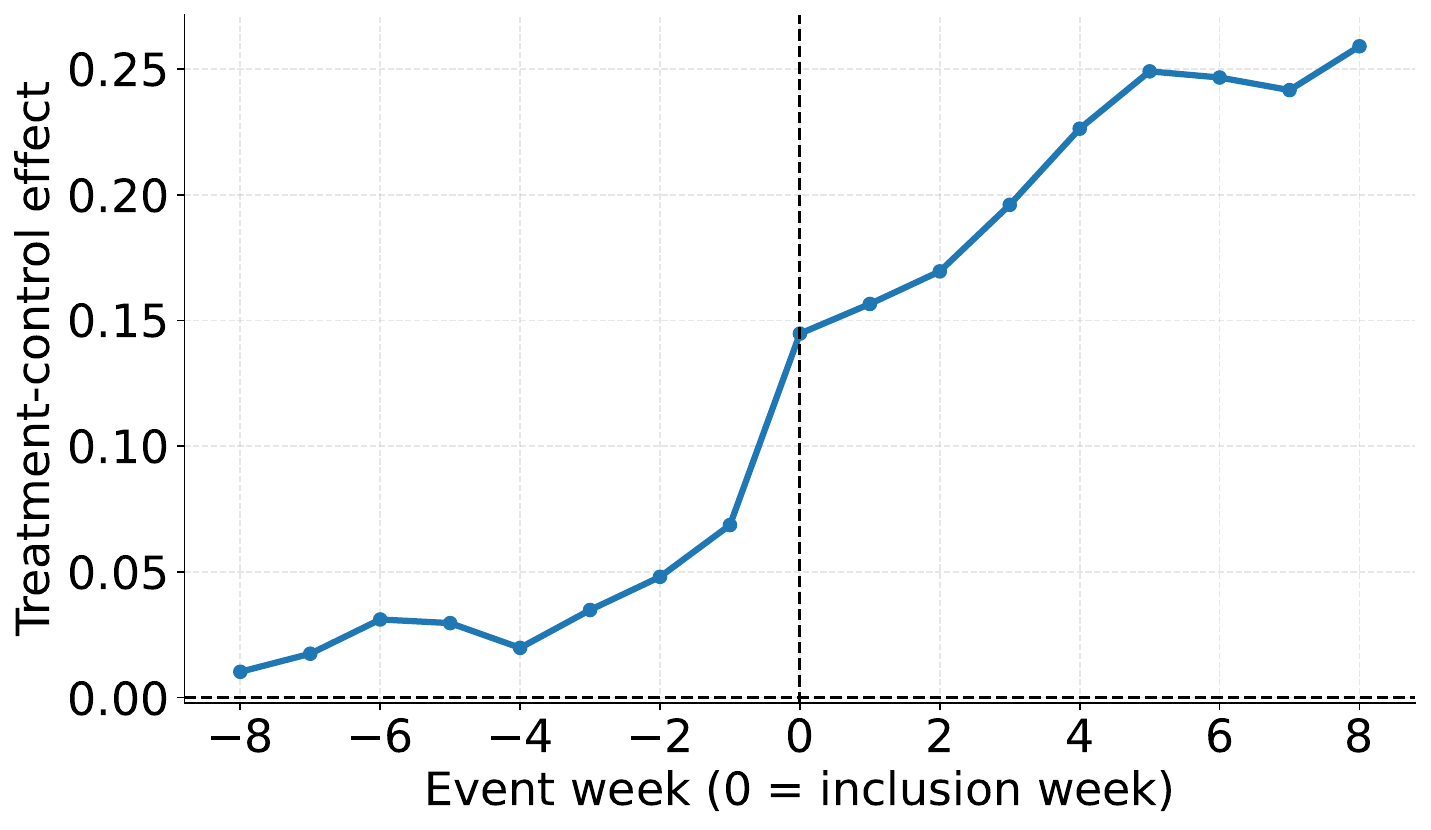}
        \caption{}
        \label{fig:pretrend-avg}
    \end{subfigure}

\caption{Event-study estimates of the diffusion impact of starter pack inclusion for the reposted once (a) and average reposts (b) criteria. 
The lines describe treatment-control differences in repost outcomes by event week relative to the inclusion week ($t=0$). 
The dashed vertical line indicates the week of first inclusion in a starter pack. 
}
    \label{fig:combined3}
\end{figure}

\noindent We compare topic classes on network density, reciprocity, and the fraction of edges formed after pack creation.
Across all three measures, the distributions largely overlap across topics, with similar central tendencies and dispersion (Fig.~\ref{fig:follownets}). 
Consistent with this visual similarity, effect sizes are small for density ($\eta^2=0.048$) and reciprocity ($\eta^2=0.024$), indicating that topic explains only a limited share of the observed variation in these structural properties. 
In contrast, the fraction of edges formed after pack creation exhibits a larger effect ($\eta^2=0.134$).
In particular we notice that \textit{Science and Research} has overall higher fraction of edges established after pack creation, an effect we attribute to the well-documented scholarly migration to Bluesky~\cite{mallapaty2024scientists}.
While the internal structure of these networks appears largely consistent across topics, their effectiveness as discovery mechanisms may still vary. 
To investigate this aspect, we examine the extent to which users adopt starter packs by following their members.
Fig.~\ref{fig:adoption} shows the median number of detected adopters per starter pack across topics. 
Adoption intensity varies across topics. 
Packs related to \textit{Ukraine and Russia}, \textit{Science and Research}, and \textit{News and Journalism} exhibit the highest median adoption levels, suggesting that informational communities are more likely to trigger adoption. 
These results are coherent with previous work showing widespread use of Bluesky related to the Russo-Ukrainian conflict~\cite{failla2024m}, as well as scholarly adoption~\cite{mallapaty2024scientists}.
In contrast, entertainment-oriented topics such as \textit{Comedy and Laughter}, \textit{Cats}, and \textit{Content Creators} show substantially lower median adoption.
The total number of detected adopters per topic largely reflects differences in the number of packs. 
In particular, the \textit{People} category dominates in absolute terms, accumulating more than 3M detected adopters overall, despite relatively modest adoption per pack.
\\ \ \\
\noindent\textbf{Diffusion Impact Analysis.}
While the previous analyses focus on structure and adoption, we now turn to the impact of starter packs on content diffusion. The input data comprise more than 227M posts and 82M repost events. 
After aggregation, the matched event-time panel contains 13.7M author–week observations, including 3.65M treated and 1.35M never-treated control rows. 
Matching uses five nearest neighbors within exact \emph{follower-count} deciles, where follower counts are measured strictly before inclusion (treated) or from the control pre-period anchor week. 
All treated authors are matched.
Prior to inclusion in a starter pack, treated authors already exhibit higher diffusion than their matched controls. In the pre-treatment window, treated authors receive on average 0.250 reposts per post compared with 0.208 for matched controls, and the probability that a post receives at least one repost is 8.89\% for treated authors versus 7.20\% for controls. These differences correspond to treated–control gaps of +0.0422 reposts per post and +1.69 percentage points in repost probability.
Following inclusion, treated authors experience a substantial improvement in repost performance relative to the matched control group. 
In the post-treatment window, treated authors receive on average 0.415 reposts per post compared with 0.208 for controls, while the probability that a post receives at least one repost rises to 10.60\% for treated authors versus 7.12\% for controls. This corresponds to an increase of roughly 1.66× in reposts per post relative to the authors' own pre-treatment performance and about 2.00× more reposts per post compared with matched controls. Similarly, the likelihood that a post receives at least one repost is about 1.49× higher than matched controls.
\\
Combining these estimates in a difference-in-differences comparison yields treatment effects of +0.165 reposts per post and +1.79 percentage points in the probability that a post receives at least one repost within seven days. 
Event-time estimates show modest positive pre-treatment leads (e.g., +0.0153 for repost incidence and +0.0273 for average reposts at $t \le -2$), indicating some upward pre-trends before inclusion (Fig.~\ref{fig:combined3}).
These pre-treatment differences suggest a partial deviation from the parallel trends assumption, consistent with the possibility that authors selected into starter packs are already on upward diffusion trajectories prior to inclusion.
However, the magnitude of these pre-trends is small relative to the notable increase observed at the time of inclusion and the subsequent post-treatment effects. 
One plausible explanation is selection: accounts added to starter packs may be added precisely \textit{because they are getting more popular}. 
Nevertheless, the largest changes in repost outcomes occur at the treatment boundary and in the post-treatment period, suggesting that inclusion in a starter pack is, at a minimum, very strongly associated with an additional improvement in short-run post diffusion beyond pre-existing trends.

\section{Conclusion}
In this paper, we provide a large-scale empirical characterization of Bluesky starter packs, examining their structure, content, and impact on content diffusion. Using a dataset of more than 50K English-language starter packs and hundreds of thousands of users, we analyze how these curated lists organize communities and influence visibility on the platform.

Our findings reveal that the starter pack ecosystem forms a highly interconnected discovery infrastructure rather than a collection of independent curated lists: packs broadly overlap, follower networks among members are dense and reciprocal, and the majority of ties predate pack creation, suggesting that packs primarily surface pre-existing communities rather than new ones.  Topically, the landscape is dominated by auto-generated personal packs, with informational communities such as science, journalism, and Ukraine-Russia discussion attracting the highest adoption rates, while entertainment-oriented packs show more diffuse engagement.  Inclusion in a starter pack is strongly associated with increased content diffusion, with treated authors experiencing roughly twice as many reposts per post as matched controls after inclusion.

Our study is subject to several limitations. The matched event-study design mitigates but cannot fully eliminate selection confounds. Aalthough matching on baseline audience size improves comparability, unobserved differences in growth trajectories, prominence, and platform-level recommendation exposure may still confound the estimated effects. Our adoption heuristic, based on bursts of simultaneous follow events, is a lower bound and likely misses cases where adopters already follow a subset of pack members. Additionally, our analysis is restricted to English-language packs, and dynamics in other linguistic communities may differ. 
Future work could extend this analysis to multilingual contexts, investigate the role of specific feed generators bundled within packs, and examine longer-term outcomes beyond the seven-day diffusion window studied here.  Moreover, the activity of highly prolific starter pack creators may warrant closer examination to better understand the characteristics of the content they produce and how it relates to the lists they curate.  Finally, future work could investigate if and how curated discovery mechanisms are influenced by user blocking behavior and moderation practices. As Bluesky continues to grow and the AT Protocol ecosystem matures, starter packs provide a natural setting to study how curated discovery mechanisms shape network structure, community formation, and information diffusion at scale.

%
%
%
 \bibliographystyle{splncs04}
 \bibliography{mybibliography}

@article{failla2024m,
  title={“I’m in the bluesky tonight”: insights from a year worth of social data},
  author={Failla, Andrea and Rossetti, Giulio},
  journal={PloS one},
  volume={19},
  number={11},
  pages={e0310330},
  year={2024},
  publisher={Public Library of Science San Francisco, CA USA}
}

@article{serrano2009extracting,
  title={Extracting the multiscale backbone of complex weighted networks},
  author={Serrano, M {\'A}ngeles and Bogun{\'a}, Mari{\'a}n and Vespignani, Alessandro},
  journal={Proceedings of the national academy of sciences},
  volume={106},
  number={16},
  pages={6483--6488},
  year={2009},
  publisher={National Academy of Sciences}
}

@article{gomes2022network,
  title={On network backbone extraction for modeling online collective behavior},
  author={Gomes Ferreira, Carlos Henrique and Murai, Fabricio and Silva, Ana PC and Trevisan, Martino and Vassio, Luca and Drago, Idilio and Mellia, Marco and Almeida, Jussara M},
  journal={Plos one},
  volume={17},
  number={9},
  pages={e0274218},
  year={2022},
  publisher={Public Library of Science San Francisco, CA USA}
}

@inproceedings{balduf2025bootstrapping,
  title={Bootstrapping social networks: Lessons from Bluesky starter packs},
  author={Balduf, Leonhard and Sokoto, Saidu and Baronchelli, Andrea and Castro, Ignacio and Kr{\'o}l, Micha{\l} and Tyson, Gareth and Pavlou, George and Scheuermann, Bj{\"o}rn and Ascigil, Onur},
  booktitle={Proceedings of the International AAAI Conference on Web and Social Media},
  volume={19},
  pages={178--192},
  year={2025}
}

@article{mcinnes2018umap,
  title={Umap: Uniform manifold approximation and projection for dimension reduction},
  author={McInnes, Leland and Healy, John and Melville, James},
  journal={arXiv preprint arXiv:1802.03426},
  year={2018}
}

@article{mcinnes2017hdbscan,
  title={hdbscan: Hierarchical density based clustering.},
  author={McInnes, Leland and Healy, John and Astels, Steve and others},
  journal={J. Open Source Softw.},
  volume={2},
  number={11},
  pages={205},
  year={2017}
}

@article{yassin2023evaluation,
  title={An evaluation tool for backbone extraction techniques in weighted complex networks},
  author={Yassin, Ali and Haidar, Abbas and Cherifi, Hocine and Seba, Hamida and Togni, Olivier},
  journal={Scientific Reports},
  volume={13},
  number={1},
  pages={17000},
  year={2023},
  publisher={Nature Publishing Group UK London}
}

@inproceedings{soliman2019characterization,
  title={A characterization of political communities on reddit},
  author={Soliman, Ahmed and Hafer, Jan and Lemmerich, Florian},
  booktitle={Proceedings of the 30th ACM conference on hypertext and Social Media},
  pages={259--263},
  year={2019}
}

@article{kim2024does,
  title={How does stress experienced on Instagram differ from threads? Comparing social media fatigue based on platform types},
  author={Kim, Sieun and Ma, Ilhwan and Son, Jeyoung},
  journal={Computers in Human Behavior},
  volume={157},
  pages={108249},
  year={2024},
  publisher={Elsevier}
}

@misc{linkedinBuildingLargeScale,
	author = {},
	title = {{B}uilding a {L}arge-{S}cale {R}ecommendation {S}ystem: {P}eople {Y}ou {M}ay {K}now --- linkedin.com},
	howpublished = {\url{https://shorturl.at/FNGgU}},
	year = {},
	note = {[Accessed 10-03-2026]},
}

@inproceedings{liben2003link,
  title={The link prediction problem for social networks},
  author={Liben-Nowell, David and Kleinberg, Jon},
  booktitle={Proceedings of the twelfth international conference on Information and knowledge management},
  pages={556--559},
  year={2003}
}

@inproceedings{gupta2013wtf,
  title={Wtf: The who to follow service at twitter},
  author={Gupta, Pankaj and Goel, Ashish and Lin, Jimmy and Sharma, Aneesh and Wang, Dong and Zadeh, Reza},
  booktitle={Proceedings of the 22nd international conference on World Wide Web},
  pages={505--514},
  year={2013}
}

@inproceedings{kywe2012survey,
  title={A survey of recommender systems in twitter},
  author={Kywe, Su Mon and Lim, Ee-Peng and Zhu, Feida},
  booktitle={International Conference on Social Informatics},
  pages={420--433},
  year={2012},
  organization={Springer}
}

@inproceedings{kang2012using,
  title={Using lists to measure homophily on twitter},
  author={Kang, Jeon Hyung and Lerman, Kristina},
  booktitle={AAAI workshop on Intelligent techniques for web personalization and recommendation},

  year={2012}
}

@article{sharif2013review,
  title={A review on search and discovery mechanisms in social networks},
  author={Sharif, Shabnam Hassanzadeh and Mahmazi, Shabnam and Navimipour, Nima Jafari and Aghdam, Behzad Farid},
  journal={International Journal of Information Engineering and Electronic Business},
  volume={5},
  number={6},
  pages={64},
  year={2013},
  publisher={Modern Education and Computer Science Press}
}

@inproceedings{oubalahcen2023recommender,
  title={Recommender systems for social networks: A short review},
  author={Oubalahcen, Houda and El Ouadghiri, Moulay Driss},
  booktitle={Proceedings of the 6th International Conference on Networking, Intelligent Systems \& Security},
  pages={1--7},
  year={2023}
}

@inproceedings{zhang2024emergence,
  title={The emergence of threads: The birth of a new social network},
  author={Zhang, Peixian and He, Yupeng and Haq, Ehsan-Ul and He, Jiahui and Tyson, Gareth},
  booktitle={International Conference on Advances in Social Networks Analysis and Mining},
  pages={69--78},
  year={2024},
  organization={Springer}
}

@inproceedings{kleppmann2024bluesky,
  title={Bluesky and the at protocol: Usable decentralized social media},
  author={Kleppmann, Martin and Frazee, Paul and Gold, Jake and Graber, Jay and Holmgren, Daniel and Ivy, Devin and Johnson, Jeromy and Newbold, Bryan and Volpert, Jaz},
  booktitle={Proceedings of the ACM Conext-2024 Workshop on the Decentralization of the Internet},
  pages={1--7},
  year={2024}
}

@article{quelle2025bluesky,
  title={Bluesky: Network topology, polarization, and algorithmic curation},
  author={Quelle, Dorian and Bovet, Alexandre},
  journal={PloS one},
  volume={20},
  number={2},
  pages={e0318034},
  year={2025},
  publisher={Public Library of Science}
}

@misc{Bluesky_2024_sp, url={https://bsky.social/about/blog/06-26-2024-starter-packs},  title={Introducing Bluesky Starter Packs},
journal={Bluesky}, year={2024}}

@article{smith2026blue,
  title={A blue start: a large-scale pairwise and higher-order social network dataset},
  author={Smith, Alyssa Hasegawa and Amburg, Ilya and Kumar, Sagar and Welles, Brooke Foucault and Landry, Nicholas W},
  journal={Scientific Data},
  year={2026},
  publisher={Nature Publishing Group UK London}
}

@article{achiam2023gpt,
  title={Gpt-4 technical report},
  author={Achiam, Josh and Adler, Steven and Agarwal, Sandhini and Ahmad, Lama and Akkaya, Ilge and Aleman, Florencia Leoni and Almeida, Diogo and Altenschmidt, Janko and Altman, Sam and Anadkat, Shyamal and others},
  journal={arXiv preprint arXiv:2303.08774},
  year={2023}
}

@article{grootendorst2022bertopic,
  title={BERTopic: Neural topic modeling with a class-based TF-IDF procedure},
  author={Grootendorst, Maarten},
  journal={arXiv preprint arXiv:2203.05794},
  year={2022}
}

@misc{mallapaty2024scientists,
  title={WHY SCIENTISTS ARE JOINING THE RUSH TO BLUESKY},
  author={Mallapaty, Smriti},
  year={2024},
  publisher={NATURE PORTFOLIO HEIDELBERGER PLATZ 3, BERLIN, 14197, GERMANY}
}

\end{document}